\documentclass{epl}
\usepackage{amssymb}

\usepackage{epsfig}

\begin{document}

\title{Will jams get worse when slow cars move over?}
\shorttitle{Will jams get worse ...}
\author{B.~Schmittmann, J.~Krometis, and R.~K.~P.~Zia}
\institute{Center for Stochastic Processes in Science and Engineering and\\
Department of Physics, Virginia Tech, Blacksburg, VA 24061-0435 USA}
\date{January 29, 2005}

\maketitle

\begin{abstract}
Motivated by an analogy with traffic, we simulate two species of particles
(`vehicles'), moving stochastically in opposite directions on 
a two-lane ring road. 
Each species prefers one lane over the other, controlled
by a parameter $0 \leq b \leq 1$ such that $b=0$ corresponds to random 
lane choice and $b=1$ to perfect `laning'. We find that the 
system displays one large cluster (`jam')
whose size increases with $b$, contrary to intuition. Even more
remarkably, the lane `charge' (a measure for the number of particles in
their preferred lane) exhibits a region of negative response: even though
vehicles experience a stronger preference for the `right' lane, 
more of them find themselves in the `wrong' one! For $b$ very close to $1$, 
a sharp transition restores a homogeneous state. Various characteristics 
of the system are computed analytically, in good agreement with
simulation data.
\end{abstract}

%\twocolumn[\hsize\textwidth
%\columnwidth\hsize\csname
%@twocolumnfalse\endcsname

\vspace{-2mm}

Driven diffusive systems \cite{Katz,reviews} have been widely studied since
they are amongst the simplest models which settle into nontrivial \emph{%
nonequilibrium} steady states. These are of fundamental interest to
physicists in their quest to formulate a theoretical framework for
nonequilibrium behavior, on a par with Gibbs ensemble theory. Moreover, such
models serve to describe numerous scientific or engineering situations
involving \emph{net} currents of mass and/or energy. Examples include
colloids sedimenting under gravity \cite{colloids}, molecular motors moving
along a microtubule \cite{motors}, or traffic flowing on a highway \cite%
{traffic}.

The behavior of such systems becomes especially interesting when two (or
more) different components (`species') move preferentially in different
directions, such as, e.g., positive and negative charges in a uniform
electric field. In the traffic analogy, the two species of particles can be
interpreted as `fast cars' and `slow trucks', viewed from a co-moving
frame. In the simplest model, cars and trucks may pass each other with a
small rate and change lanes randomly. For ring roads with two or more lanes
and no exits, two distinct phases are observed: on one side of the phase
boundary, typical configurations are disordered and particles move freely;
on the other side, configurations are spatially inhomogeneous and jammed up %
\cite{KornissLL,Korniss2L}. In particular, Monte Carlo simulations for
two-lane ($L \times 2$) roads show one large \emph{macroscopic} jam, of size 
$O(L)$, containing almost all vehicles. Only a few particles
(`travellers') are found outside the jam, having just escaped by repeated
passing.

We should note that recent analytic \cite{1dexactsol} and simulational \cite%
{GSZ} studies suggest that this jam is `merely' a finite-size effect:
As $L\rightarrow \infty $, the system consists of a distribution of
jam-sizes controlled by a finite length $\ell _{0}$. However, $\ell _{0}$
may be as large as $10^{70}$\cite{1dexactsol}, rendering the study of
systems with physically accessible sizes of great interest, as is our focus
here.

It is natural to ask whether the jams will get shorter when drivers develop
a \emph{lane preference}. For simplicity, we explore this question for a
two-lane periodic system, modeling cars/trucks preferring the `fast/slow' lane. We
introduce a parameter, $b$, for the bias towards the preferred lane. Thus, $%
b=0$ denotes no preference (random lane changes) while $b=1$ corresponds to
perfect lane segregation (`laning') of cars and trucks. With no vehicles in
the `wrong' lane, there will be no jams for $b=1$. As $b$ increases from
zero, we may expect the number of vehicles in the `wrong' lane to decrease, 
creating fewer `obstacles' for those in the `right' lane, and leading to 
smaller jams. In fact, we observe just the opposite: As $b$ increases, the
cluster \emph{grows} in size until a sharp first order transition, very
close to $b=1$, restores the system to homogeneity. Even more remarkably,
the number of vehicles in their preferred lane depends \emph{%
non-monotonically} on $b$, exhibiting a region of \emph{negative} response.
Clearly, the system is more complex than one might have anticipated.

This letter is organized as follows. We first describe our model and a few
technical details of the simulations. We then present our data, followed by
supporting analytic arguments. We conclude with a summary and open
questions. More details will be reported elsewhere \cite{Krometis}.

Our model is defined on a $L\times 2$ lattice with periodic boundary
conditions in the $x$-direction. With $x=0,1,...,L-1$ and $y=0,1$, each site 
$(x,y)$ can be empty or occupied by a positive ($+$) or a negative ($-$) particle. 
Configurations are specified in terms of an occupation variable $q(x,y)$, 
taking the values $0$ or $\pm 1$, depending on the `charge' at the site. 
For later convenience, we define the local `mass' by $m(x,y)\equiv \left|
q(x,y)\right| $, the positive particle density by $p\left( x,y\right) \equiv 
\frac{1}{2}\left[ m\left( x,y\right) +q\left( x,y\right) \right] $ and
similarly, the negative case by $n\equiv \frac{1}{2}\left[ m-q\right] $.
The system is neutral and half-filled; $N\equiv L/2$ is the number of
positive particles in the system. Our rates are motivated by the traffic
analogy. Positive (negative) particles never move left (right). The allowed
moves $+0\rightarrow 0+$ and $0-\rightarrow -0$ occur with rate $1$, and
`passing' $+-\rightarrow -+$ takes place with rate $\gamma <1$. Turning to
lane changes, a positive (negative) particle in lane $0$ ($1$) hops
into a hole in lane $1$ ($0$) with rate $1$, while the opposite move
occurs with rate $1-b$; finally, particles can exchange positions across
lanes with rate $\gamma $ or $\gamma (1-b)$, depending on whether the move
takes them into their preferred lanes or not.

\begin{figure}[!t]
\begin{center}
\epsfig{file=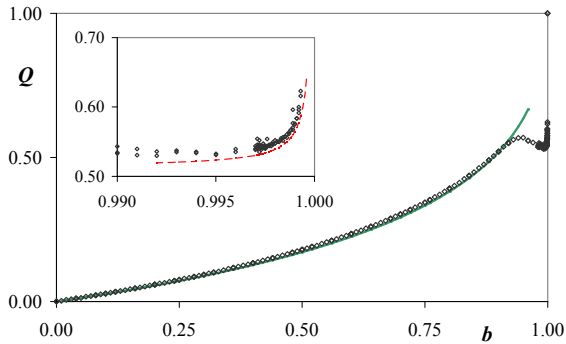, width=3in} \vspace{-3mm} 
\caption{Charge $Q(b)$ vs $b$ for $L=10^{3}$ and $\gamma=0.1$. 
The MC data are shown as open
diamonds; the region of negative response is observed at $0.95 \leq b \leq
0.99$. The solid and dashed lines follow from a mean-field theory (see
text). The inset magnifies the large $b$ data.}
\vspace{-3mm}
\label{fig1}
\end{center}
\end{figure}

The updating rule is random sequential. We choose a nearest-neighbor pair (bond) 
at random and attempt to exchange its contents. One Monte Carlo step (MCS)
consists of selecting $L$ bonds. Simulations lasted from $5\times 10^{5}$ up
to $10^{7}$ MCS. Typically, $2\times 10^{5}$ MCS were discarded to ensure
that the steady state was reached, and data were then taken every $100$ MCS.
Unless otherwise stated, all measured quantities are time averages in the
steady state, denoted by $\left\langle \cdot \right\rangle $. Here, we focus
on data for $\gamma=0.1$ and $L=10^{3}$; other $L$'s, ranging from $100$ to 
$10^{4}$ have also been simulated. All show the same \emph{qualitative} 
behaviors. Quantitatively, we find only very weak finite-size effects 
(below $5 \%$) for sizes $250 \leq L \leq 10^{4}$, giving us confidence in our findings 
over a significant range of $L$.

For $b=0$, ordered states display one large cluster. To distinguish such
configurations from disordered ones, we define an order parameter, $\Omega
(b)\equiv \left\langle \left( 1/2\right) \sin \left( \pi /L\right) \left|
\sum_{x,y}e^{2\pi ix/L}m(x,y)\right| \right\rangle $ which is $O(L^{-1/2})$
if particles are distributed randomly, and takes its maximum value, 1, for a
single jam containing all particles. The measured value $\Omega (0)=0.940$
indicates that a few particles (the `travellers') remain outside the
jam. Remarkably, $\Omega (b)$ increases with $b$, demonstrating that the
cluster actually \emph{grows} in size. This pattern persists up to $%
b=0.998\pm 0.001$ where $\Omega $ drops abruptly (from $0.972$ to $0.045$)
as the system returns to disorder. This sharp transition appears to be first
order, since it displays hysteresis \cite{Krometis}.

As soon as $b>0$, particles begin to favor one lane over the other, leading
to a `charge' imbalance. Clearly, it suffices to monitor the \emph{total
charge} in, e.g., lane $1$, $Q(b)\equiv (2/L)\sum_{x}\left\langle
q(x,1)\right\rangle $, normalized such that $0\leq Q(b)\leq 1$. Shown in
Fig.~1, $Q(b)$ first increases monotonically, as one would expect. At $%
b=0.95 $, however, $Q(b)$ displays a maximum and enters a region of \emph{%
negative response}, i.e., $dQ/db < 0$: More vehicles are found in the
`wrong' lanes, even though their preference for the `right' lane has
increased! Since the order parameter remains perfectly monotonic here, this
behavior must be associated with a restructuring of the jam \emph{interior}.
Near $b\simeq 0.99$, $Q(b)$ goes through a minimum (inset) and becomes
`normal' again. Finally, it jumps abruptly to the disordered value ($\simeq
1 $), correlated with the drop in $\Omega(b) $.

A detailed view of the jammed state is provided by measuring steady-state
mass and charge profiles of each lane, $\left\langle m(x,y)\right\rangle $
and $\left\langle q(x,y)\right\rangle $. Since the jams form randomly and
diffuse slowly during a run, we shift their centers of mass to $x=0$ \emph{%
before} averaging, by tracking the \emph{phase} of the Fourier transformed
mass density.

\begin{figure}[!t]
\begin{center}
\epsfig{file=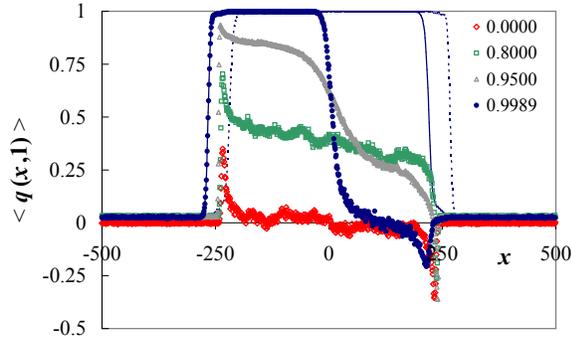, width=3in} \vspace{-3mm} 
\caption{Charge profiles $\left\langle q(x,1)\right\rangle $ vs $x$ for
different $b$; $L=10^{3}$ and $\gamma=0.1$. 
For comparison, the solid (dashed) lines are the
mass profiles $\left\langle m(x,y)\right\rangle $ of lane $1$ ($0$), for $%
b=0.9989$. }
\label{fig2}
\vspace{-3mm}
\end{center}
\end{figure}

The disordered state is spatially uniform. In contrast, profiles in the
jammed phase are highly nontrivial and depend strongly on $b$. Due to an
underlying symmetry, it suffices to focus our attention on just one lane,
say, lane $1$ (preferred by positive charges). Fig.~2 shows four charge
profiles $\left\langle q(x,1)\right\rangle $, at different values of $b$.
Focusing first on common features, each profile displays two distinct
regions. Outside the large cluster, there are few travellers. Denoted by $%
m^{\ast }$ and $q^{\ast }$, the mass and charge densities are uniform.
Inside the cluster, positive/negative charges tend to accumulate at its
left/right edge, so that $\left\langle q(x,1)\right\rangle $ decreases
from left to right. The mass profiles (plotted only for $b=0.9989$) show
essentially no holes in the interior for any $b$. At the jam edges, $%
\left\langle q(x,1)\right\rangle $ varies rapidly, reminiscent of a shock.
Beyond these global features, remarkable differences emerge. For $b=0$, the
traveller region is neutral in each lane, and $\left\langle
q(x,1)\right\rangle $ is odd with respect to the center of mass, $x=0$. As $b
$ increases, the traveller region becomes charged, reaching a maximum of $%
q^{\ast }(b)\simeq 0.032$ at $b=0.81$. The remainder of the profile looks
similar to its $b=0$ counterpart, but shifted upwards by a constant as
positive particles begin to favor this lane. As $b$ approaches $0.95$, i.e.,
the \emph{maximum} of $Q(b)$ (cf. Fig.~1), the profile changes shape
dramatically\emph{\ inside} the cluster, while the local charge \emph{outside%
} is now \emph{lower}: $q^{\ast }(0.95)\simeq 0.029$. The region of negative
response in $Q$ is associated with this `reshaping' of the interior
profile. As $b$ increases further, beyond $b\simeq 0.99$, each lane develops
an extended block which is (almost) entirely occupied by its preferred
species. As we will see below, these \emph{solid} domains act as blockages
through which no currents can flow. The remainder of the cluster is still 
\emph{mixed}, containing both positive and negative particles with a
nontrivial charge gradient. At the jam edges, another new feature emerges
here, namely, the shock in lane $0$ is distinctly \emph{offset} from its
counterpart in lane $1$. As illustrated in Fig.~2 for $b=0.9989$, the left
edge of the cluster in lane $1$ is located $47\pm 1$ lattice sites further
left than its counterpart in lane $0$! The traveller region is now almost
perfectly charge-segregated, with $m^{\ast }(0.9989)\simeq q^{\ast
}(0.9989)\simeq 0.026$. For profiles of this type, $Q_{1}(b)$ is, once
again, monotonically increasing. Eventually, at $b=0.9990$, the large jam
disintegrates and the system reverts to disorder.

A remarkable feature of this system is the presence of highly \emph{%
nonuniform} current densities, both in the longitudinal (in-lane) as well as
the transverse (cross-lane) directions. Focusing on just the \emph{positive}
particles, say, we let $J^{+}(x-1,x;y)$ denote the (\emph{average}) in-lane
current from site $x-1$ to $x$ in lane $y$, and $j^{+}(x)$ denote the
cross-lane current from lane $0$ into lane $1$, at $x$. Thus, $j^{+}>0$
indicates a net current into the preferred lane. In the steady state,
particle conservation ensures that $%
J^{+}(x-1,x;y)-J^{+}(x,x+1;y)=(-1)^{y}j^{+}(x)$ for all $(x,y)$. In the
extreme cases with $b\cong 1$ (e.g., $0.9989$), particles in preferred lanes
form solid domains making up half of the jam. Since bonds occupied by two
identical particles carry no current, the in-lane current is essentially zero
through these solid domains. Yet, the traveller region clearly carries a 
\emph{nonzero }current. As a result, particles outside the jam flow mainly
in their preferred lane, and then jump -- via a suppressed transition! --
into the `wrong lane' so as to circumvent the solid blockage. Such highly
suppressed moves are facilitated by the offset of the cluster boundaries in
the two lanes. In the `wrong lane,' particles move through the mixed region
of the cluster, and return to the preferred lane near the center of the
cluster. This behavior is understandable, since a particle in the `wrong
lane' faces a solid block of the opposite species. By crossing back into the
`right lane,' they make their way through a mixed region and exit the jam.
This phenomenon is well illustrated in Fig.~3, in which the solid
circles show $j^{+}<0$ in a sizable region around $x=-250$, followed
by a positive peak near $x=0$. As a contrast, for $b\lesssim 0.8$ cases,
where such solid blockages have not formed yet, the currents are quite
different (open squares): Positive charges still jump into the
unfavored lane near the left edge of the cluster, but the associated peak in 
$j^{+}$ is much narrower. Since they face no solid domains in either lane,
there is no need to return to the favored lane until the end of the jam. The
analogue in circuits for this two-lane jam would be two resistors in
parallel, with the incoming current split appropriately between them at one
end, only to rejoin at the other end. In all other parts of the system, the
cross-lane currents are statistically indistinguishable from zero, so that
the $J^{+}$'s are essentially uniform (within such regions). 

\begin{figure}[!t]
\begin{center}
\epsfig{file=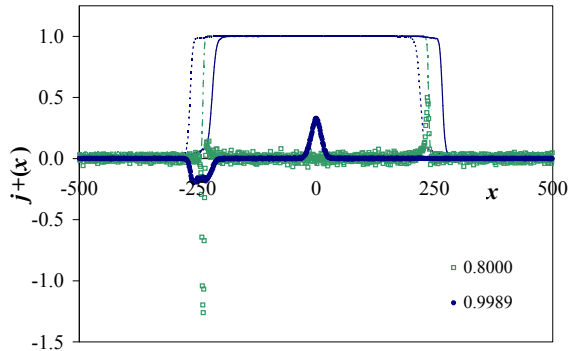, width=3in} \vspace{-3mm} 
\caption{Cross-lane currents $j^{+}(x)$ for two values of $b$. The solid
(dashed) line is the mass profile of lane $1$ ($0$) for $b=0.9989$; the
dashed-dotted line is the mass profile of lane $1$ (indistinguishable from
lane $0$) for $b=0.8000$. }
\vspace{-3mm}
\label{fig3}
\end{center}
\end{figure}

In the remainder of this letter, we present selected analytic results for
the steady state, with details to be reported elsewhere \cite{Krometis}.
Specifically, we aim to compute $Q(b)$ over a range of $b$ within a simple
mean-field theory, using some input from the simulations. Omitting the $%
\left\langle \cdot \right\rangle $ brackets from now on, we let $p(x,y)$ and 
$n(x,y)\ $denote the configurational averages of the local densities of
positive and negative particles, with $m(x,y)$ and $q(x,y)$ defined similarly. 
With the center of mass still at $x=0$,
symmetry dictates $p(x,0)=n(-x,1)$\ and $p(x,1)=n(-x,0)$. Outside the jam,
the densities are uniform, to be denoted by $p^{\ast }(y)$ and $n^{\ast }(y)$%
, with $p^{\ast }(0)=$ $n^{\ast }(1)$ and $p^{\ast }(1)=$ $n^{\ast }(0)$.
The traveller mass densities in each lane are equal: $p^{\ast }(0)+n^{\ast
}(0)=p^{\ast }(1)+n^{\ast }(1)\equiv m^{\ast }$, while $p^{\ast }(1)-n^{\ast
}(1)=n^{\ast }(0)-p^{\ast }(0)\equiv q^{\ast }$ denotes the traveller
charge. The interior of the cluster contains essentially no holes whence $%
m(x,y)\simeq 1$ there. As a result, for each value of $b$, we need to
determine only two independent densities, say, $m^{\ast }$ and $q^{\ast }$
to characterize the traveller region, and a \emph{single function}, e.g., $%
p(x,0)$, to characterize the cluster. Of course, the traveller region and
the cluster are coupled to one another, through currents flowing along the
lanes and from one lane to the other. In mean-field theory, the net \emph{%
in-lane} current of positive particles, from site $x-1$ to $x$, can be
written as $J^{+}(x-1,x;y)=p(x-1,y)\left[ 1-m(x,y)\right] +\gamma
p(x-1,y)n(x,y)$. Here, the first term reflects the move of a positive
particle from $(x-1,y)$ into an empty site at $(x,y)$, and the second term
reflects the exchange of a positive and a negative particle. Similarly, we
can write the net \emph{cross-lane} current of positive particles, from lane 
$0$ into lane $1$, as $j^{+}(x)=p(x,0)\left[ 1-m(x,1)\right]
-(1-b)p(x,1)\left[ 1-m(x,0)\right] +\gamma p(x,0)n(x,1)-\gamma
(1-b)p(x,1)n(x,0)$ with all other currents following by symmetry. In
particular, when summed over both lanes, the net mass current vanishes, and
the net charge current is constant, independent of $x$.

Considering the traveller region first, having constant densities there
implies $0=j^{+}(x)$ which is consistent with the current data (Fig.~3).
This allows us to compute all densities in the traveller region in terms of
the single parameter $m^{\ast }$. The resulting expressions simplify for
small $\gamma $; we find, e.g., $q^{\ast }=bm^{\ast }/(2-b)+O(\gamma )$.
Both $q^{\ast }$ and $m^{\ast }$ are easily measured so that this relation
can be tested: It holds remarkably well over the full range of $b$ \cite%
{Krometis}, giving us confidence in our mean-field theory.

Next, we turn to the interior of the cluster. The expressions for the
currents simplify considerably here, due to the absence of holes. For $b\leq
0.9$, the data suggest $j^{+}(x)\simeq 0$ inside the cluster. Exploiting
this observation for $x=0$ and invoking symmetries, we can compute the local
charge at the center of the cluster: $q(0,1)=(1-\sqrt{1-b})^{2}/b$.
Moreover, for $b\leq 0.9$, $q(x,1)-q(0,1)$ is, to good approximation, an 
\emph{odd} function of $x$ inside the jam. If we assume, for simplicity,
that the travellers can be neglected so that the cluster extends over L/2
sites, we may express the total charge in lane $1$
as $Q(b)= (2/L) \sum_{x=-L/2}^{L/2}q(x,1)\simeq (2/L) q(0,1)$ which is
plotted in Fig.~1 as a solid line. Clearly, it follows the data remarkably
well, up to $b\leq 0.9$.

Next, we focus on profiles just before the onset of disorder. For $b\gtrsim
0.99$, the traveller region is essentially charge-segregated, i.e.~$p^{\ast
}(1)\simeq n^{\ast }(0)\simeq m^{\ast }$. 
As a result, the $+$ current there is given by $%
J^{+}(x-1,x;y)\simeq m^{\ast }\left( 1-m^{\ast }\right) $. When these
particles encounter the solid block of positive charge which forms the left
edge of the cluster, they change lanes and enter the mixed region of the
cluster in lane $0$. Since there is no leakage current into the other lane
here, the mixed region can be viewed as a single species asymmetric
exclusion process in the maximal current phase, whence $J^{+}(x-1,x;y)=%
\gamma /4$, up to finite size corrections. Equating these two
currents yields $m^{\ast }=(1-\sqrt{1-\gamma }%
)/2=0.026$ for $\gamma =0.1$, in excellent agreement with the data. Next, we
consider the cross-lane current which is localized in the offset region.
Here, correlations turn out to be essential: If a positive particle has just
performed a \emph{suppressed} jump from lane $1$ into $0$, it leaves a hole
behind. If this bond is selected again, before either particle or hole have
moved away, the particle will return to its original location, and no
contribution to the current has resulted. If we approximate the number of 
\emph{suppressed }jumps over the length $\Delta $ of the offset by its
mean-field form $\Delta (1-b)(1-m^{\ast })$, the total net cross-lane
current is simply the fraction $\alpha $ of such moves which are not
`undone', i.e., $\alpha \Delta (1-b)(1-m^{\ast })$. We have not been able
to find a good analytic estimate for $\alpha $, but it is easily measured in
simulations: $\alpha =1-N_{f}/N_{s}$, where $N_{f}$ ($N_{s}$) is the number
of favored (suppressed) moves between the lanes in the offset region during
a run. For example, we measure $\alpha =0.503\pm 0.003$ for $b=0.9989$.
Equating currents again gives $\Delta =m^{\ast }/\alpha (1-b)$. Thus, we
predict $\Delta =46.3$, for $b=0.9989$ and $\gamma=0.1$, 
in excellent agreement with the
simulation result $\Delta =$ $47\pm 1$. Similar arguments \cite{Krometis}
yield the lane charge of such profiles as $Q(b)=1/(2-2m^{\ast })+\Delta /L$.
Assuming $\alpha \simeq 1/2$, this expression compares very well with the
data for $b\geq 0.995$ (dashed line in the inset of Fig.~1).

To summarize, we consider two species of particles (`vehicles'), biased to
move in opposite directions along a two-lane road. Vehicles can pass each
other (with a small rate $\gamma $) and change lanes, controlled by a
parameter $b$. We find that (\emph{i}) for all $b<0.9996$, the system
displays one large cluster (`jam') whose size increases monotonically with 
$b$. This is quite remarkable, given that a larger $b$ \emph{reduces} the
number of obstacles which a particle encounters as it travels down its
preferred lane. For $b$ very close to unity, a first order transition
restores homogeneity. (\emph{ii}) The lane `charge' $Q$ exhibits a region of
negative response for $0.950\leq b\leq 0.992$. This is arguably our most
surprising result: even though the particles experience a stronger
preference for their own lane, more of them find themselves in the `wrong'
lane. Preliminary data for systems with different numbers of positive
and negative particles show that this behavior persists for a significant
range of system `charge'. 
(\emph{iii}) The system exhibits nontrivial mass current loops:
particles change into the unfavored and back into the favored lane at
well-defined locations in the system. This observation, in conjunction with
particle conservation, allows us to compute the lane charge analytically, in
good agreement with the data \emph{except} in the regime displaying negative
response. Here, a very complex restructuring of the cluster interior takes
place, requiring further analysis \cite{Krometis}.

\emph{Acknowledgements.} We thank J.T. Mettetal and I.T. Georgiev for
stimulating discussions. This research is supported in part by the US
National Science Foundation through grants DMR-0088451 and DMR-0414122.

\end{document}